\def\aa{{A\&A}}

\def\apj{{ApJ}}

\def\mnras{{MNRAS}}
\def\nat{{Nature}}

\documentclass[cup5b]{caps}
\usepackage{graphicx}
\usepackage{amssymb}
\usepackage{ociwsymp3e}  

\HeadText{A. J. R. Sanderson and T. J. Ponman}  

\def\myplottwo#1#2{\centering \hspace*{-1.4cm}\leavevmode
\includegraphics[width=.38\columnwidth,angle=270]{#1} \hfil
\includegraphics[width=.38\columnwidth,angle=270]{#2}}

\begin{document}

\pagenumbering{arabic}

\author[]{A. J. R. Sanderson$^{1,2}$ and T. J. Ponman$^{1}$
\\
(1) University of Birmingham, Birmingham, UK\\
(2) The University of Illinois at Urbana-Champaign, IL, USA}

\chapter{The X-ray Scaling Properties of Virialized Systems}

\begin{abstract}

Virialized systems, such as clusters and groups of galaxies, represent an
ideal laboratory for investigating the formation and evolution of structure
on the largest scales. Furthermore, the properties of the gaseous
intracluster medium provide key insights into the influence of important
non-gravitational processes like energy injection and radiative cooling, 
resulting from feedback associated with star formation, for example.

We have assembled a very large X-ray sample of virialized systems, spanning
over two decades in halo mass. Each object has high-quality X-ray data
available, enabling a full deprojection analysis to be made; for a
subsample we have additionally determined the deprojected optical light
distribution.

We find clear evidence of a departure from the simple expectations of
self-similarity. The intracluster medium is more spatially extended and
systematically less dense in smaller haloes, and there is evidence of an
entropy excess in the hot gas. Our results favour a significant role for
both non-gravitational heating and radiative cooling in modifying the
properties of this gas, although we find no clear evidence of significantly 
enhanced star formation efficiency in groups.

\end{abstract}

\section{Introduction}
Simple models of the collapse and subsequent formation of virialized
systems predict that such objects will exhibit self-similarity in their
behaviour (e.g. Navarro et al. 1995). Shock heating of the intracluster
gas, driven by gravitational infall, establishes the X-ray properties of
these systems, such that they scale simply with the total mass of the
halo. This leads to the expectation of a constant, universal gas fraction 
in virialized systems. Additionally, assuming haloes form from material of
fixed density, a number of well defined scaling relations can be derived.

However, it is now clear that the presence of additional physics is
required to provide a more complete picture of the state of the gas. The
well studied relation between X-ray luminosity and mean temperature
exhibits a logarithmic slope steeper than expected ($L_{\mathrm{X}}
\propto T^2$) in clusters (e.g. Edge \& Stewart 1991; Arnaud \& Evrard 1999). 
Further steepening is observed at group scales (e.g. Helsdon \& Ponman
2000).  Recent work has focused on the role of non-gravitational heating
(e.g. Babul et al. 1999; Tozzi \& Norman 2001) and radiative cooling
(e.g. Muanwong et al. 2001; Voit \& Bryan 2001) of the gas in explaining
the observed departures from self-similarity. 

However, existing observations have been unable to discriminate between
these competing theories, since both heating and cooling of gas can give
rise to a reduction in central density and a suppression of X-ray
luminosity in the central regions of cooler systems. In this short paper we
present an overview of an ongoing project to study a large sample of
virialized systems, in order to establish their scaling properties and
address the issue of self-similarity breaking and its causes.

Throughout this paper we assume a Hubble constant of
70~~km~s$^{-1}$~Mpc$^{-1}$.

\section{The sample}
Our sample incorporates rich clusters of galaxies, through groups and
included two galaxy-sized haloes, comprising 66 objects in total.  For each
system we have fitted analytical profiles to describe the deprojected
density and temperature of the intracluster medium (ICM) as a function of
radius.  We have excluded from our sample those systems with obvious
evidence of X-ray substructure, where the assumption of hydrostatic
equilibrium is not reasonable. The redshift range spanned is $z=$
0.0036--0.208 (0.035 median).

Our analysis combines three existing samples, based on \textit{ROSAT} and 
\textit{ASCA} data, and a detailed description of our method can be found
in Sanderson et al. (2003). With these data we are able to determine the
gravitating mass profile, virial radius and other derived quantities in a
self-consistent fashion. For a subset of 32 groups and clusters, we have
also measured the distribution of the galaxies themselves (using the APM
catalogue), allowing us to calculate the stellar mass contribution
(Sanderson \& Ponman 2003b). We convert optical light to stellar
mass, assuming a mass-to-light ratio for early-type galaxies of 7~$h_{70}$
Solar in the $B$ band (Pizzella et al. 1997).

\section{Results}

\subsection{Gas fraction \& mass-to-light ratio}
    \begin{figure}[t]
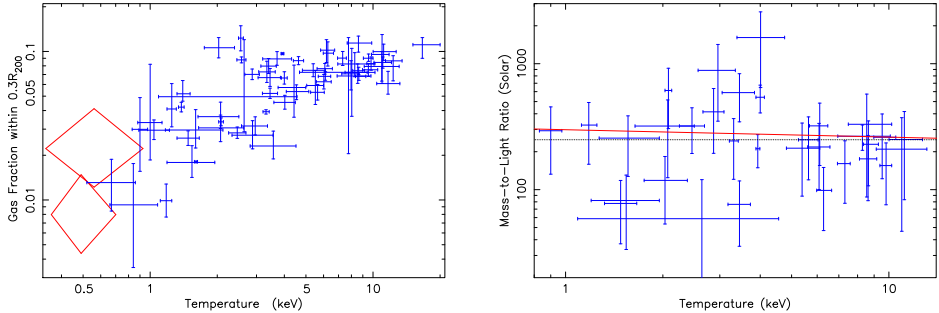

    \myplottwo{fig1a.eps}{fig1b.eps}
    \caption{\textit{Left}: Gas fraction within 0.3R$_{200}$ as a function 
       of mean X-ray temperature; the diamonds represent the two galaxies.
       \textit{Right}: $B_{\mathrm{j}}$ band mass-to-light ratio versus
        temperature. See text for details.}
    \label{fig:fgas_MLR}
  \end{figure}

There is a strong (6$\sigma$) trend in gas fraction within 0.3R$_{200}$
with system temperature (left panel, Figure~\ref{fig:fgas_MLR}), implying
that groups and cool clusters are not self-similar compared to rich
clusters. This depletion of gas in cooler systems could be due to:
\begin{itemize}
\item Preheating: the energetically boosted ICM is progressively more weakly 
 captured by the less massive haloes of cooler systems
\item Radiative cooling: gas drops out to form stars; a process which is 
 more efficient in the cooler \& denser ICM of groups.
\end{itemize}
From our optical analysis we have derived the $B_{\mathrm{j}}$ band
mass-to-light ratio for a subset of 32 groups and clusters, which is
plotted against X-ray temperature in the right panel of
Figure~\ref{fig:fgas_MLR}. The red solid line represents the best fitting
power law, which has a logarithmic slope of $-0.06\pm0.17$.  The black
dotted line shows the median value for the sample, of 249\,$h_{70}$
$(M/L)_{\odot}$. It is clear that there is no significant trend in $M/L$,
indicating that star formation efficiency does not vary substantially with
halo mass.

\subsection{Spatial mass distribution}
\begin{figure}
    \hspace*{-1.4cm}
    \includegraphics[width=5.3cm,angle=270]{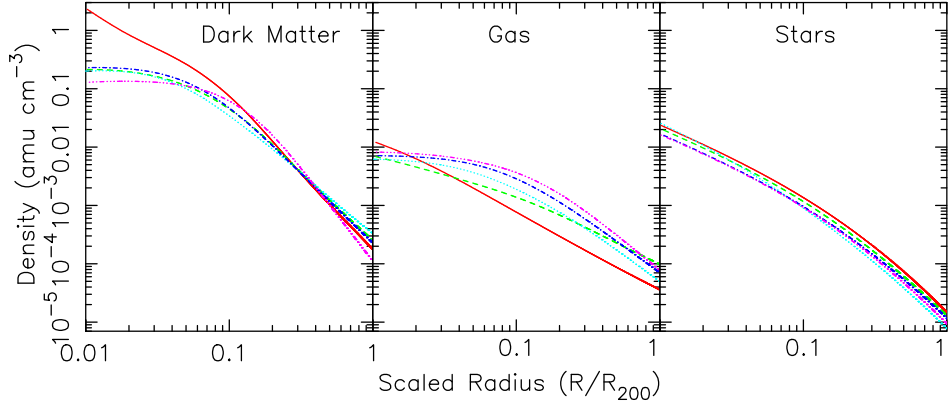}
    \caption{Density profiles for the cluster mass components, grouped by 
      temperature. The red solid line represents the coolest systems
      (0.3--2.0~keV), increasing in temperature through green dashed 
      (2.0--2.9~keV), cyan dotted (2.9--4.6~keV), blue dot-dashed 
      (4.6-8.0~keV) and magenta dot-dot-dot-dashed (8.0--17~keV). Gas 
      profiles are for the full X-ray sample (66 objects); stellar \& dark 
      matter profiles are for the optical sub-sample (32 objects).}
    \label{fig:rho(r)}
  \end{figure}

The X-ray data enable both the gas and total mass density profiles to be
determined. Our optical sub-sample yields the stellar baryon contribution,
and therefore allows the dark matter distribution to be inferred. Density
profiles for the individual cluster mass components are plotted in
Figure~\ref{fig:rho(r)} as a function of scaled radius (i.e. normalized to 
our nominal virial radius, $R_{200}$). The profiles have been grouped into
five temperature bands for clarity, with approximately the same number of
profiles in each band.

The dark matter is the most self-similar component, showing no systematic
trend with temperature, but for an enhancement in central concentration
within $\sim0.1R_{200}$ for the coolest temperature band (0.3--2.0~keV).
The intracluster gas is clearly not self-similar and shows a significant
decrease in density with decreasing temperature. The stellar distribution
shows a surprisingly self-similar shape across the sample, with a small
spread in normalization; there is some evidence for a mild enhancement in 
the stellar density in the two coolest bands.

Comparison between the three panels in Figure~\ref{fig:rho(r)} reveals that 
the gas is the most spatially extended mass component, followed by the stars
and dark matter, in agreement with the findings of David et al. (1995).

\subsection{Gas entropy}
Since entropy is conserved in any adiabatic process, it serves as an
excellent probe of non-gravitational physics, increasing in proportion to
the virial temperature of a cluster for the self-similar case. It is
defined, for convenience, as
\begin{equation}
S=T/n_\mathrm{e}^{2/3}~\textrm{keV} \textrm{cm}^{2} ,
\end{equation}
where $n_\mathrm{e}$ is the electron number density of the gas and $T$ is its
temperature.

\begin{figure}
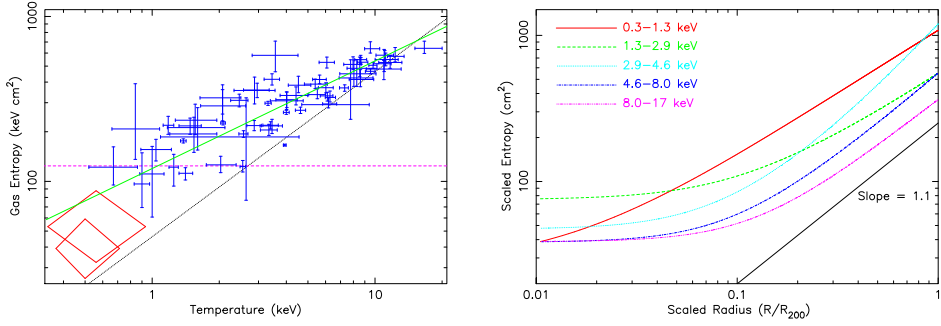

    \myplottwo{fig3a.eps}{fig3b.eps}
    \caption{\textit{Left}: Gas entropy at 0.1R$_{200}$ versus temperature;
     the diamonds represent the two galaxies. \textit{Right}: Gas entropy 
     profiles, each scaled by $1/T$, grouped into bands according to the 
     mean system temperature ($T$). See text for details.}
    \label{fig:entropy}
  \end{figure}

The entropy at a fiducial radius of 0.1R$_{200}$ is plotted against system
temperature in the left panel of Figure~\ref{fig:entropy}. The solid green
line is the best fit to the data, which has a logarithmic slope of
$0.65\pm0.05$. This relation is significantly shallower than the
self-similar slope of unity, normalized to the mean of the hottest 8
clusters and plotted as the solid black line. The entropy `floor' (Ponman
et al. 1999) value of 124 $h_{70}^{-1/3}$~keV cm$^2$ from Lloyd-Davies et
al. (2000) is shown as the magenta horizontal line, but it is not clear if
this represents a true lower limit to the entropy in galaxy groups, since
the overall trend is flatter than $S\propto T$ for all temperatures. In
general, the data appear to be more consistent with radiative cooling
models (e.g. Voit \& Bryan 2001) than preheating ones, which predict
$S\propto T$ for more massive clusters.

The right panel of Figure~\ref{fig:entropy} shows entropy (scaled by mean
temperature) as a function of radius for the sample, grouped into five
temperature bands to suppress scatter. The shape of the profiles is broadly
similar, but they are shifted in normalization such that the cooler systems
posses a higher scaled entropy. However, there is no evidence for a large
isentropic core in any of the profiles, which would be expected if the
extra entropy was injected by some non-gravitational heating mechanism
(e.g. Balogh et al. 1999). These findings are confirmed by a recent
analysis of two galaxy groups using \textit{XMM-Newton}; Mushotzky et
al. (2003) find no evidence for an entropy floor within $\sim0.3R_{200}$,
which would otherwise be expected if their hot gas had been preheated prior
to virialization.

Furthermore, it is clear from the right panel of Figure~\ref{fig:entropy}
that the excess entropy is present \emph{at all radii}, and not just in the
central regions.  However, for all the profiles, the entropy in the
outskirts of the halo appears consistent with the expectations of shock
heating of accreting gas, which lead to a logarithmic slope of 1.1 (Tozzi
\& Norman 2001), as shown by the black solid line in the right panel of
Figure~\ref{fig:entropy}.

The entropy properties of the sample are examined more thoroughly in Ponman 
et al. (2003).

\section{Conclusions}
We have studied the scaling properties of a large sample of virialized
systems, spanning a wide range of mass. We find clear evidence of a departure
from self-similarity in the properties of the hot gas in these systems: Groups
have substantially lower gas fractions than clusters, and there is evidence
of excess entropy in the intracluster medium on all mass scales. Our entropy
results are inconsistent with the predictions of preheating models, and 
provide a better match to models incorporating the effects of radiative
cooling and feedback from subsequent star formation. 

However, we do not find significant variation in mass-to-light ratio with
mass, implying that star formation efficiency is not greatly enhanced in
groups. This demonstrates that radiative cooling cannot be the only mechanism
responsible for the departure from self-similarity; non-gravitational 
heating of the gas must also have had some influence.

We are grateful to Alexis Finoguenov, Ed Lloyd-Davies and Maxim Markevitch
for providing X-ray data and contributing to the original analysis.

\end{document}